\documentclass{llncs}
\usepackage[latin9]{inputenc}
\usepackage{array}
\usepackage{verbatim}
\usepackage{float}
\usepackage{amsbsy}
\usepackage{amstext}
\usepackage{graphicx}
\usepackage[unicode=true]
 {hyperref}
\usepackage{breakurl}

\makeatletter

\providecommand{\tabularnewline}{\\}
\floatstyle{ruled}
\newfloat{algorithm}{tbp}{loa}
\providecommand{\algorithmname}{Algorithm}
\floatname{algorithm}{\protect\algorithmname}

\usepackage{fancyhdr}
\@ifundefined{showcaptionsetup}{}{%
 \PassOptionsToPackage{caption=false}{subfig}}
\usepackage{subfig}
\makeatother

\begin{document}

\title{Generic Black-Box End-to-End Attack Against State of the Art API
Call Based Malware Classifiers}

\author{Ishai~Rosenberg, Asaf~Shabtai, Lior~Rokach, and Yuval~Elovici}

\institute{Software and Information Systems Engineering Department, Ben Gurion
University}

\institute{ishairos@post.bgu.ac.il}
\maketitle
\begin{abstract}
In this paper, we present a black-box attack against API call based
machine learning malware classifiers, focusing on generating adversarial
sequences combining API calls and static features (e.g., printable
strings) that will be misclassified by the classifier without affecting
the malware functionality. We show that this attack is effective against
many classifiers due to the transferability principle between RNN
variants, feed forward DNNs, and traditional machine learning classifiers
such as SVM. We also implement GADGET, a software framework to convert
any malware binary to a binary undetected by malware classifiers,
using the proposed attack, without access to the malware source code.
\end{abstract}

\section{Introduction}

Machine learning malware classifiers, in which the model is trained
on features extracted from the analyzed file, have two main advantages
over current signature based/black list classifiers: 1) Automatically
training the classifier on new malware samples saves time and expense,
compared to manually analyzing new malware variants. 2) Generalization
to currently unseen and unsigned threats is better when the classifier
is based on features and not on a fingerprint of a specific and exact
file (e.g., a file's hash).

Next generation anti-malware products, such as \href{https://www.cylance.com/en_us/products/our-products/protect.html}{Cylance},
\href{https://www.crowdstrike.com/wp-content/brochures/Falcon-Prevent-FINAL.pdf}{CrowdStrike},
and \href{https://www.sophos.com/en-us/medialibrary/PDFs/factsheets/sophos-intercept-x-dsna.pdf}{Sophos},
use machine and deep learning models instead of signatures and heuristics.
Those models can be evaded and in this paper, we demonstrate an evasive
\emph{end-to-end attack}, generating a malware binary that can be
executed while not being detected by such machine learning malware
classifiers.

Application programming interface (API) calls, often used to characterize
the behavior of a program, are a common input choice for a classifier
and used by products such as \href{https://www.sentinelone.com/insights/endpoint-protection-platform-datasheet/}{SentinelOne}.
Since only the sequence of API calls gives each API call its context
and proper meaning, API call sequence based classifiers provide state
of the art detection performance (\cite{Huang16}).

Machine learning classifiers and algorithms are vulnerable to different
kinds of attacks aimed at undermining the classifier's integrity,
availability, etc. One such attack is based on the generation of adversarial
examples which are originally correctly classified inputs that are
perturbed (modified) so they (incorrectly) get assigned a different
label. In this paper, we demonstrate an attack like this on binary
classifiers that are used to differentiate between malicious and benign
API call sequences. In our case, the adversarial example is a malicious
API call sequence, originally correctly classified, which is classified
by the classifier as benign (a form of evasion attack) after the perturbation
(which does not affect the malware functionality).

Generating adversarial examples for API sequences differs from generating
adversarial examples for images \cite{Carlini17}, which is the main
focus of the existing research, in two respects: 1) API sequences
consist of discrete symbols with variable lengths, while images are
represented as matrices with fixed dimensions, and the values of the
matrices are continuous. 2) In adversarial API sequences one must
verify that the original functionality of the malware remains intact.
Attacks against RNN variants exist (\cite{Hu17b,Papernot16d}), but
they are not practical attacks, in that they don't verify the functionality
of the modified samples or handle API call arguments and non-sequence
features, etc. The differences from our attack are specified in Section
2.

The contributions of our paper are as follows:
\begin{enumerate}
\item We implement a novel \emph{end-to-end black-box method} to generate
adversarial examples for many state of the art machine learning malware
classifiers. This is the first attack to be evaluated against RNN
variants (like LSTM), feed forward DNNs, and traditional machine learning
classifiers (such as SVM). We test our implementation on a large dataset
of 500,000 malware and benign samples.
\item Unlike previous papers that focus on images, we focus on the cyber
security domain. We implement GADGET, an evasion framework generating
a new malware binary with the perturbed features \emph{without access
to the malware source code} that allows us to \emph{verify that the
malicious functionality remains intact}.
\item Unlike previous papers, we extend our attack to \emph{bypass multi-feature
(e.g., static and dynamic features) based malware classifiers}, to
fit real world scenarios.
\item We focus on \emph{the principle} of \emph{transferability} in RNN
variants. To the best of our knowledge, this is \emph{the first time
it has been evaluated in the context of RNNs and in the cyber security
domain}, proving that the proposed attack is effective against the
largest number of classifiers ever reviewed in a single study: RNN,
LSTM, GRU, and their bidirectional and deep variants, and feed forward
DNN, 1D CNN, SVM, random forest, logistic regression, GBDT, etc.
\end{enumerate}

\section{Background and Related Work}

Most black-box attacks rely on the concept of \emph{adversarial example
transferability} \cite{Szegedy14}: Adversarial examples crafted against
one model are also likely to be effective against other models, even
when the models are trained on different datasets. This means that
the adversary can train a \emph{surrogate model}, which has decision
boundaries similar to the original model, and perform a white-box
attack on it. Adversarial examples that successfully fool the surrogate
model are likely to fool the original model as well \cite{Papernot16c}.
A different approach uses the confidence score of the targeted DNN
to estimate its gradients directly instead of using the surrogate
model's gradients to generate adversarial examples \cite{Chen17}.
However, attacker knowledge of confidence scores (not required by
our attack) is unlikely in black-box scenarios. \emph{Decision based
attack,} which uses only the target classifier's classes, without
the confidence score, result in lower attack effectiveness and higher
overhead \cite{Rosenberg17b}. 

In \emph{mimicry attacks}, an attacker is able to code a malicious
exploit that mimics the system calls' trace of benign code, thus evading
detection \cite{Wagner02}. Several methods were presented: 1) \emph{Disguise
attacks} - Causing benign system calls to generate malicious behavior
by modifying only the system calls' parameters. 2) \emph{No-op Attacks}
- Adding semantic \emph{no-ops} - system calls with no effect, or
those with an irrelevant effect, e.g., opening a non-existent file.
3) \emph{Equivalence attack} - Using a different system call sequence
to achieve the same (malicious) effect.

The search for adversarial examples can be formalized as a minimization
problem \cite{Szegedy14}:

\begin{equation}
\arg_{\boldsymbol{\mathbf{r}}}\min f(\boldsymbol{\mathbf{x}}+\boldsymbol{\mathbf{r}})\neq f(\boldsymbol{\mathbf{x}})\:s.t.\:\boldsymbol{\mathbf{x}}+\boldsymbol{\mathbf{r}}\in\boldsymbol{\mathbf{D}}
\end{equation}
The input \textbf{$\boldsymbol{\mathbf{x}}$}, correctly classified
by the classifier $f$, is perturbed with \textbf{$\boldsymbol{\mathbf{r}}$}
such that the resulting adversarial example \textbf{$\boldsymbol{\mathbf{x}}+\boldsymbol{\mathbf{r}}$}
remains in the input domain \textbf{$\boldsymbol{\mathbf{D}}$}, but
is assigned a different label than \textbf{$\mathbf{\boldsymbol{x}}$}.

A substitute model was trained with inputs generated by augmenting
the initial set of representative inputs with their FGSM \cite{Goodfellow15}
perturbed variants, and then the substitute model was used to craft
adversarial samples \cite{Papernot16c}. This differs from our paper
in that: 1) It deals \emph{only} with convolutional neural networks,
as opposed to all state of the art classifiers, including RNN variants.
2) It deals with images and doesn't fit the attack requirements of
the cyber security domain, i.e., not harming the malware functionality.
3) No end-to-end framework to implement the attack in the cyber-security
domain was presented.

A white-box evasion technique for an Android static analysis malware
classifier was implemented using the gradients to find the element
whose addition would cause the maximum change in the benign score,
and add this feature to the adversarial example \cite{Grosse16}.
In contrast to our work, this paper didn't deal with RNNs or dynamic
features which are more challenging to add without harming the malware
functionality. This study also did not focus on a generic attack that
can affect many types of classifiers, as we do. Finally, our black-box
assumption is more feasible than a white-box assumption. In Section
5.3 we created a black-box variant of this attack.

API call uni-grams were used as static features, as well \cite{Hu17a}.
A generative adversarial network (GAN) was trained to generate adversarial
samples that would be classified as benign by the discriminator which
uses labels from the black-box model. This attack doesn't fit sequence
based malware classifiers (LSTM, etc.). In addition, the paper does
not present a end-to-end framework which preserves the code's functionality.
Finally, GANs are known for their unstable training process \cite{Arjovsky17},
making such an attack method hard to rely on.

A white-box adversarial example attack against RNNs, demonstrated
against LSTM architecture, for sentiment classification of a movie
reviews dataset was shown in \cite{Papernot16d}. The adversary iterates
over the movie review's words \textbf{$\boldsymbol{x}[i]$} in the
review and modifies it as follows:

\begin{equation}
\boldsymbol{\mathbf{x}}[i]=\arg\min_{\boldsymbol{\mathbf{z}}}||sign(\boldsymbol{\mathbf{x}}[i]-\boldsymbol{\mathbf{z}})-sign(J_{f}(\boldsymbol{\mathbf{x}})[i,f(\boldsymbol{\mathbf{x}})])||\:s.t.\:\boldsymbol{\mathbf{z}}\in\boldsymbol{\mathbf{\textrm{D}}}
\end{equation}

where $f(\boldsymbol{\mathbf{x}})$ is the original model label for
\textbf{$\boldsymbol{\mathbf{x}}$}, and $J_{f}(\boldsymbol{\mathbf{x}})[i,j]=\frac{\partial f_{j}}{\partial x_{i}}(\boldsymbol{\mathbf{x}})$.
This differs from our paper in that: 1) We present a black-box attack,
not a white-box attack. 2) We implement a practical cyber domain attack.
For instance, we don't modify existing API calls, because while such
an attack is relevant for reviews - it might damage a malware functionality
which we wish to avoid. 3) We deal with multiple-feature classifiers,
as in real world malware classifiers. 4) Our attack has better performance,
as shown in Section 4.3.

Concurrently and independently from our work, a RNN GAN to generate
invalid APIs and insert them into the original API sequences was proposed
\cite{Hu17b}. Gumbel-Softmax, a one-hot continuous distribution estimator,
was used to deliver gradient information between the generative RNN
and the substitute RNN. Null APIs were added, but while they were
omitted to make the generated adversarial sequence shorter, they remained
in the gradient calculation of the loss function. This decreases the
attack effectiveness compared to our method (88\% vs. 99.99\% using
our method, for an LSTM classifier). In contrast, our attack method
doesn't have this difference between the substitute model and the
black-box model, and our generated API sequences are shorter. This
also makes our adversarial example faster. Unlike \cite{Hu17b}, which
only focused on LSTM variants, we also show our attack's effectiveness
against other RNN variants such as GRUs and conventional RNNs, bidirectional
and deep variants, and non-RNN classifiers (including both feed forward
networks and traditional machine learning classifiers such as SVM),
making it truly generic. Moreover, the usage of Gumbel-Softmax approximation
in \cite{Hu17b} makes this attack limited to one-hot encoded inputs,
while in our attack, any word embedding can be used, making it more
generic. In addition, the stability issues associated with GAN training
\cite{Arjovsky17}, which might not converge for specific datasets,
apply to the attack method mentioned in \cite{Hu17b} as well, making
it hard to rely on. While such issues might not be visible when using
a small dataset (180 samples in \cite{Hu17b}), they become more apparent
when using larger datasets like ours (500,000 samples). Finally, we
developed an end-to-end framework, generating a mimicry attack (Section
5). While previous works inject arbitrary API call sequences that
might harm the malware functionality (e.g., by inserting the \emph{ExitProcess()}
API call in the middle of the malware code), our attack modifies the
code such that the original functionality of the malware is preserved
(Section 5.1). Moreover, our approach works in real world scenarios
including hybrid classifiers/multiple feature types (Section 5.3)
and API arguments (Section 5.2), non of which is addressed by \cite{Hu17b}.

\section{Methodology}

\subsection{Black-Box API Call Based Malware Classifier}

Our classifier's input is a sequence of API calls made by the inspected
code. In this section, it uses only the API call type and not its
arguments or return value. IDSs that verify the arguments tend to
be much slower (4-10 times slower, in \cite{Tandon06}). One might
claim that considering arguments would make our attack easier to detect.
This could be done, e.g., by looking for irregularities in the arguments
of the API calls (e.g., invalid file handles, etc.) or by considering
only successful API calls and ignoring failed APIs. In order to address
this issue, we don't use null arguments that would fail the function.
Instead, arguments that are valid but do nothing, such as writing
into a temporary file instead of an invalid file handle, are used
in our framework, as described in Section 5. We also discuss an extension
of our attack that handles API call arguments in Section 5.2.

Since API call sequences can be long (some samples in our dataset
have millions of API calls), it is impossible to train on the entire
sequence at once due to GPU memory and training time constraints.
Thus, we used a sliding window approach: Each API call sequence is
divided into windows with size $m$. Detection is performed on each
window in turn, and if any window is classified as malicious, the
entire sequence is malicious. This method helps detect cases like
malicious payloads injected into goodware (e.g., using Metasploit),
where only a small subset of the sequence is malicious. We use one-hot
encoding for each API call type in order to cope with the limitations
of sklearn's implementation of decision trees and random forests,
as mentioned in \href{https://roamanalytics.com/2016/10/28/are-categorical-variables-getting-lost-in-your-random-forests/}{https://roamanalytics.com/2016/10/28/are-categorical-variables-getting-lost-in-your-random-forests/}.
The output of each classifier is binary (is the inspected code malicious
or not). The tested classifiers and their hyper parameters are described
in Section 4.2.

\subsection{Black-Box API Call Based Malware Classifier Attack}

The proposed attack has two phases: 1) creating a surrogate model
using the target classifier as a black-box model, and 2) generating
adversarial examples with white-box access to the surrogate model
and using them against the attacked black-box model, by the transferability
property.

\subsubsection{Creating a Surrogate Model}

We use Jacobian-based dataset augmentation, an approach similar to
\cite{Papernot16c}. The method is specified in Algorithm 1.

We query the black-box model with synthetic inputs selected by a Jacobian-based
heuristic to build a surrogate model $\hat{f}$, approximating the
black-box model $f$\textquoteright s decision boundaries. While the
adversary is unaware of the architecture of the black-box model, we
assume the basic features used (the recorded API call types) are known
to the attacker. In order to learn decision boundaries similar to
the black-box model while minimizing the number of black-box model
queries, the synthetic training inputs are based on prioritizing directions
in which the model\textquoteright s output varies. This is done by
evaluating the sign of the Jacobian matrix dimension corresponding
to the label assigned to input \textbf{$\boldsymbol{\mathbf{x}}$}
by the black-box model, $sign(J_{\hat{f}}(\boldsymbol{\mathbf{x}})[f(\boldsymbol{\mathbf{x}})])$,
as calculated by FGSM (\cite{Goodfellow15}). We use the Jacobian
matrix of the surrogate model, since we don't have access to the Jacobian
matrix of the black-box model. The new synthetic data point $\boldsymbol{\mathbf{x}}+\epsilon sign(J_{\hat{f}}(\boldsymbol{\mathbf{x}})[f(\boldsymbol{\mathbf{x}})])$
is added to the training set.

\begin{algorithm}[tbph]
\caption{Surrogate Model Training}
\begin{centering}
\textbf{Input}: $f$ (black-box model), T (training epochs), $X_{1}$(initial
dataset), $\epsilon$ (perturbation factor)
\par\end{centering}
Define architecture for the surrogate model $A$

\textbf{for} t=1..T:

\ \ \ \ $D_{t}=\left\{ (\boldsymbol{\mathbf{x}},f(\boldsymbol{\mathbf{x}}))|\mathbf{x}\in X_{t}\right\} $\ \ \#
Label the synthetic dataset using the black-box model

\ \ \ \ $\hat{f_{t}}=train(A,D_{t})$ \ \ \# (Re-)Train the
surrogate model

\ \ \ \ $X_{t+1}=\left\{ \boldsymbol{\mathbf{x}}+\epsilon sign(J_{\hat{f_{t}}}(\boldsymbol{\mathbf{x}})[f(\boldsymbol{\mathbf{x}})])|\mathbf{x}\in X_{t}\right\} \cup X_{t}$
\ \ \# Perform Jacobian-based dataset augmentation

\textbf{return} $\hat{f_{T}}$
\end{algorithm}

On each iteration we add a synthetic example to each existing sample.
The surrogate model dataset size is: $|X_{t}|=2^{t-1}|X_{1}|$ 

The samples used in the initial dataset, $X_{1}$, were randomly selected
from the test set distribution, but they were not included in the
training and test sets to prevent bias. $X_{1}$ should be representative
so the dataset augmentation covers all decision boundaries to increase
the augmentation's effectiveness. For example, if we only include
samples from a single family of ransomware in the initial dataset,
we will only be focusing on a specific area of the decision boundary,
and our augmentation would likely only take us in a certain direction.
However, as shown in Section 4.3, this doesn't mean that all of the
malware families in the training set must be represented to achieve
good performance.

\subsubsection{Generating Adversarial Examples}

An adversarial example is a sequence of API calls classified as malicious
by the classifier that is perturbed by the addition of API calls,
so that the modified sequence will be misclassified as benign. In
order to prevent damaging the code's functionality, we cannot remove
or modify API calls; we can only add additional API calls. In order
to add API calls in a way that doesn't hurt the code's functionality,
we generate a \emph{mimicry attack} (Section 5). Our attack is described
in Algorithm 2.

\begin{algorithm}[tbph]
\caption{Adversarial Sequence Generation}
\begin{centering}
\textbf{Input}: $f$ (black-box model), $\hat{f}$ (surrogate model),
\textbf{$\boldsymbol{\mathbf{x}}$} (malicious sequence to perturb,
of length $l$), $n$ (size of adversarial sliding window), $D$ (vocabulary)
\par\end{centering}
\textbf{for} each sliding window $\boldsymbol{\mathbf{w}}_{j}$ of
$n$ API calls in $\boldsymbol{\mathbf{x}}$:

\ \ \ \ $\boldsymbol{\boldsymbol{\mathbf{w}}_{j}}^{*}=\boldsymbol{\boldsymbol{\mathbf{w}}_{j}}$

\ \ \ \ \textbf{while} $f(\boldsymbol{\boldsymbol{\mathbf{w}}_{j}^{*}})=malicious$:

\ \ \ \ \ \ \ \ Randomly select an API's position i in $\boldsymbol{\mathbf{w}}$

\ \ \ \ \ \ \ \ \# Insert a new adversarial API in position
$i\in\{1..n\}$:

\ \ \ \ \ $\boldsymbol{\boldsymbol{\mathbf{w}}_{j}^{*}}[i]=\arg\min_{api}||sign(\boldsymbol{\boldsymbol{\mathbf{w}}_{j}}^{*}-\boldsymbol{\boldsymbol{\mathbf{w}}_{j}^{*}}[1:i-1]\perp api\perp\boldsymbol{\boldsymbol{\mathbf{w}}_{j}^{*}}[i:n-1])-sign(J_{\hat{f}}(\boldsymbol{\boldsymbol{\mathbf{w}}_{j}})[f(\boldsymbol{\boldsymbol{\boldsymbol{w}}_{j}})])||$

\ \ \ \ Replace $\boldsymbol{\mathbf{w}}_{j}$ (in $\boldsymbol{\mathbf{x}}$)
with $\boldsymbol{\boldsymbol{\mathbf{w}}_{j}}^{*}$

\textbf{return} (perturbed) $\mathbf{\boldsymbol{x}}$
\end{algorithm}

$D$ is the vocabulary of available features, that is, the API calls
recorded by the classifier. The adversarial API call sequence length
of $l$ might be different than $n$, the length of the sliding window
API call sequence that is used by the adversary. Therefore, like the
prediction, the attack is performed sequentially on $\left\lceil \frac{l}{n}\right\rceil $
windows of $n$ API calls. Note that the knowledge of $m$ (the window
size of the classifier, mentioned in Section 3.1) is not required,
as shown in Section 4.3. $\perp$ is the concatenation operation.
$\boldsymbol{\boldsymbol{\mathbf{w}}_{j}^{*}}[1:i-1]\perp api\perp\boldsymbol{\boldsymbol{\mathbf{w}}_{j}^{*}}[i:n-1]$
is the insertion of the encoded API vector in position $i$ of $\boldsymbol{\mathbf{w}}_{j}^{*}$.
The adversary randomly chooses $i$ since he/she does not have any
way to better select $i$ without incurring significant statistical
overhead. Note that an insertion of an API in position $i$ means
that the APIs from position $i..n$ ($\boldsymbol{\mathbf{w}}_{j}^{*}[i:n]$
) are ``pushed back'' one position to make room for the new API
call, in order to maintain the original sequence and preserve the
original functionality of the code. Since the sliding window has a
fixed length, the last API call, $\boldsymbol{\boldsymbol{\mathbf{w}}_{j}^{*}}[n]$,
is ``pushed out'' and removed from $\boldsymbol{\mathbf{w}}_{j}^{*}$
(this is why the term is $\perp\boldsymbol{\boldsymbol{\mathbf{w}}_{j}^{*}}[i:n-1]$,
as opposed to $\perp\boldsymbol{\boldsymbol{\mathbf{w}}_{j}^{*}}[i:n]$).
The APIs ``pushed out'' from $\boldsymbol{\mathbf{w}}_{j}$ will
become the beginning of $\boldsymbol{\mathbf{w}}_{j+1}$, so no API
is ignored.

The newly added API call is $\boldsymbol{\boldsymbol{\mathbf{w}}_{j}^{*}}[i]=\arg\min_{api}||sign(\boldsymbol{\boldsymbol{\mathbf{w}}_{j}}^{*}-\boldsymbol{\boldsymbol{\mathbf{w}}_{j}^{*}}[0:i]\perp api\perp\boldsymbol{\boldsymbol{\mathbf{w}}_{j}^{*}}[i:n-1])-sign(J_{\hat{f}}(\boldsymbol{\boldsymbol{\mathbf{w}}_{j}})[f(\boldsymbol{\boldsymbol{\boldsymbol{w}}_{j}})])||$.
$sign(J_{\hat{f}}(\boldsymbol{\boldsymbol{\mathbf{w}}_{j}})[f(\boldsymbol{\boldsymbol{\boldsymbol{w}}_{j}})])$
gives us the direction in which we have to perturb the API call sequence
in order to reduce the probability assigned to the malicious class,
$f(\boldsymbol{\mathbf{x}})$, and thus change the predicted label
of the API call sequence. However, the set of legitimate API call
embeddings is finite. Thus, we cannot set the new API to any real
value. We therefore find the API call $api$ in $D$ whose insertion
directs us closest to the direction indicated by the Jacobian as most
impactful on the model\textquoteright s prediction. We iteratively
apply this heuristic until we find an adversarial input sequence misclassified
as benign. Note that in \cite{Papernot16d} the authors \emph{replaced}
a word in a movie review, so they only needed a single element from
the Jacobian (for word $i$, which was replaced). All other words
remained the same, so no gradient change took place. In contrast,
since we \emph{add} an API call, all of the API calls following it
shift their position, so we consider the aggregated impact.

While the proposed attack is designed for API call based classifiers,
it can be generalized to any adversarial sequence generation. This
generalization is a high performance in terms of attack effectiveness
and overhead (Equations 4 and 5). This can be seen in Section 4.3,
where we compare the proposed attack to \cite{Papernot16d} for the
IMDB sentiment classification task. In Section 4.4 we show why the
same adversarial examples generated against the surrogate model would
be effective against both the black-box model and other types of classifiers
due to the principle of \emph{transferability.}

We assume that the attacker knows what API calls are available and
how each of them is encoded (one-hot encoding in this paper). This
is a commonly accepted assumption about the attacker's knowledge \cite{Huang11}.

\section{Experimental Evaluation}

\subsection{Dataset}

Our dataset contains 500,000 files (250,000 benign samples and 250,000
malware samples), including the latest variants. We have ransomware
families such as Cerber, Locky, Ramnit, Matsnu, Androm, Upatre, Delf,
Zbot, Expiro, Ipamor. and other malware types (worms, backdoors, droppers,
spyware, PUA, and viruses), each with the same number of samples,
to prevent a prediction bias towards the majority class. 80\% of the
malware families' (like the Virut virus family) samples were distributed
between the training and test sets, to determine the classifier's
ability to generalize to samples from the same family. 20\% of the
malware families (such as the WannaCry ransomware family) were used
only on the test set to assess generalization to an unseen malware
family. The temporal difference between the training set and the test
set is several months (meaning all test set samples are newer than
the training set samples), based on VirusTotal's 'first seen' date.
We labeled our dataset using \href{https://www.virustotal.com/}{VirusTotal},
an on-line scanning service which contains more than 60 different
security products. Our ground truth is that a malicious sample is
one with 15 or more positive (i.e., malware) classifications from
the 60 products. A benign sample is one with zero positive classifications.
All samples with 1-14 positives were omitted to prevent false positive
contamination of the dataset.

We ran each sample in Cuckoo Sandbox, a commonly-used malware analysis
system, for two minutes per sample.\footnote{Tracing only the first seconds of a program execution might not detect
certain malware types, like ``logic bombs'' that commence their
malicious behavior only after the program has been running some time.
However, this can be mitigated both by classifying the suspension
mechanism as malicious, if accurate, or by tracing the code operation
throughout the program execution life-time, not just when the program
starts.} We parsed the JSON file generated by Cuckoo Sandbox and extracted
the API call sequences generated by the inspected code during its
execution. The extracted API call sequences are the malware classifier's
features. Although the JSON can be used as raw input for a neural
network classifier (as done in \cite{Rosenberg17}), we parsed it,
since we wanted to focus only on API calls without adding other features,
such as connected network addresses, which are also extracted by Cuckoo
Sandbox.

The overview of the malware classification process is shown in Figure
1. Figure 2(a) present a more detailed view of the classifier's structure.

\begin{figure}[htbp]
\begin{centering}
\textsf{\includegraphics[scale=0.33]{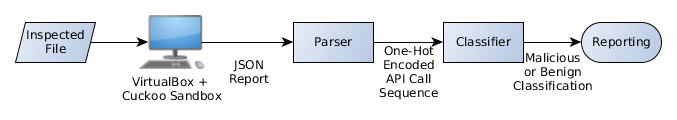}}
\par\end{centering}
\caption{Overview of the Malware Classification Process}
\end{figure}

We run the samples on a \href{https://www.virtualbox.org/}{VirtualBox}'s
snapshot with Windows 8.1 OS,\footnote{While it is true that the API calls sequence would vary across different
OSs or configurations, both the black-box classifier and the surrogate
model generalize across those differences, as they capture the ``main
features'' over the sequence, which are not vary between OSs.} since most malware target the Windows OS.

Cuckoo Sandbox is a tool known to malware writers, some of whom write
code to detect if the malware is running in a Cuckoo Sandbox (or on
virtual machines) and if so, the malware quit immediately to prevent
reversing efforts. In those cases, the file is malicious, but its
behavior recorded in Cuckoo Sandbox (its API call sequence) isn't
malicious, due to its anti-forensic capabilities. To mitigate such
contamination of our dataset, we used two countermeasures: 1) We applied
\href{https://github.com/Yara-Rules/rules}{YARA rules} to find samples
trying to detect sandbox programs such as Cuckoo Sandbox and omitted
all such samples. 2) We considered only API call sequences with more
than 15 API calls (as in \cite{Pascanu15}), omitting malware that,
e.g., detect a VM and quit. This filtering left us with about 400,000
valid samples, after balancing the benign samples number. The final
training set size is 360,000 samples, 36,000 of which serve as the
validation set. The test set size is 36,000 samples. All sets are
balanced between malicious and benign samples. One might argue that
the evasive malware that apply such anti-VM techniques are extremely
challenging and relevant. However, in this paper we focus on the adversarial
attack. This attack is generic enough to work for those evasive malware
as well, assuming that other mitigation techniques (e.g., anti-anti-VM),
would be applied.

\subsection{Malware Classifier Performance}

No open source or commercial trail versions of API calls based deep
learning intrusion detection systems are available, as such products
target enterprises. Dynamic models are not available in VirusTotal
as well. Therefore, we created our own black-box malware classifiers.
This also allows us to evaluate the attack effectiveness (Equation
4) against many classifier types. 

. We limited our maximum input sequence length to $m=140$ API calls
(longer sequence lengths, e.g., $m=1000$, had no effect on the accuracy)
and padded shorter sequences with zeros. A zero stands for a null
API in our one-hot encoding. Longer sequences are split into windows
of $m$ API calls, and each window is classified in turn. If any window
is malicious the entire sequence is considered malicious. Thus, the
input of all of the classifiers is a vector of $m=140$ API call types
in one-hot encoding, using 314 bits, since there were 314 monitored
API call types in the Cuckoo reports for our dataset. The output is
a binary classification: malicious or benign. An overview of the LSTM
architecture is shown in Figure 2(a). 

\begin{figure}
\begin{centering}
\subfloat[Dynamic Classifier Architecture]{\includegraphics[scale=0.25]{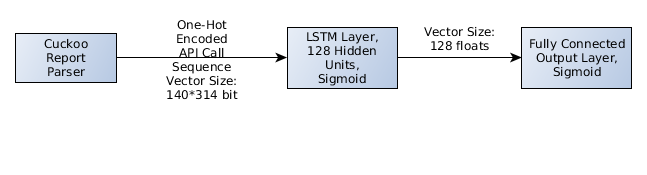}}\hfill{}\subfloat[Hybrid Classifier Architecture]{\includegraphics[scale=0.25]{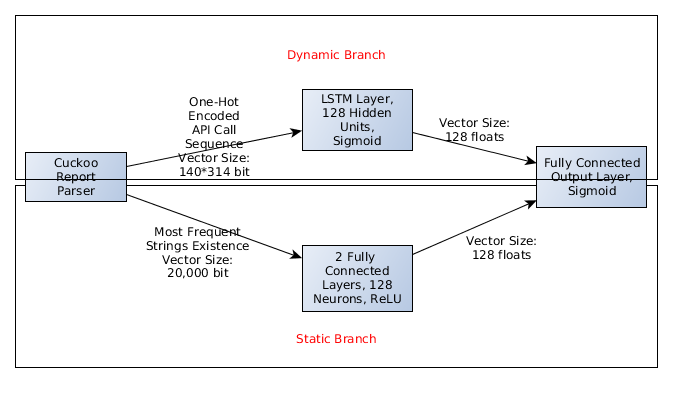}}
\par\end{centering}
\caption{Classifier Architecture Overview}
\end{figure}

We used the \href{https://keras.io/}{Keras} implementation for all
neural network classifiers, with TensorFlow used for the back end.
\href{https://github.com/dmlc/xgboost/}{XGBoost} and \href{http://scikit-learn.org/stable/}{Scikit-Learn}
were used for all other classifiers.

The loss function used for training was binary cross-entropy. We used
the Adam optimizer for all of the neural networks. The output layer
was fully-connected with sigmoid activation for all NNs. We fine-tuned
the hyper parameters for all classifiers based on the relevant state
of the art papers, e.g., window size from \cite{Pascanu15}, number
of hidden layers from \cite{Huang16} and \cite{Grosse16}, dropout
rate from \cite{Huang16}, and number of trees in a random forest
classifier and the decision tree splitting criteria from \cite{Rosenberg16}.
For neural networks, a rectified linear unit, $ReLU(x)=max(0,x)$,
was chosen as an activation function for the input and hidden layers
due to its fast convergence compared to $sigmoid()$ or $\tanh()$,
and dropout was used to improve the generalization potential of the
network. Training was conducted for a maximum of 100 epochs, but convergence
was usually reached after 15-20 epochs, depending on the type of classifier.
Batch size of 32 samples was used.

The classifiers also have the following classifier-specific hyper
parameters: DNN - Two fully-connected hidden layers of 128 neurons,
each with ReLU activation and a dropout rate of 0.2; CNN - 1D ConvNet
with 128 output filters, stride length of one, 1D convolution window
size of three and ReLU activation, followed by a global max pooling
1D layer and a fully connected layer of 128 neurons with ReLU activation
and a dropout rate of 0.2; RNN, LSTM, GRU, BRNN, BLSTM, bidirectional
GRU - a hidden layer of 128 units, with a dropout rate of 0.2 for
both inputs and recurrent states; Deep LSTM and BLSTM - Two hidden
layers of 128 units, with a dropout rate of 0.2 for both inputs and
recurrent states in both layers; Linear SVM and logistic regression
classifiers - A regularization parameter C=1.0 and L2 norm penalty;
Random forest classifier - Using 10 decision trees with unlimited
maximum depth and the Gini criteria for choosing the best split; Gradient
boosted decision tree - Up to 100 decision trees with a maximum depth
of 10 each.

We measured the performance of the classifiers using the accuracy
ratio, which applies equal weight to both FP and FN (unlike precision
or recall), thereby providing an unbiased overall performance indicator:

\begin{equation}
\mathit{accuracy}=\frac{TP+TN}{TP+FP+TN+FN}
\end{equation}

where: TP are true positives (malicious samples classified as malicious
by the black-box classifier), TN are true negatives, FP stands for
false positives (benign samples classified as malicious), and FN are
false negatives. The FP rate of the classifiers varied between 0.5-1\%.\footnote{The FP rate was chosen to be on the high end of production systems.
A lower FP rate would mean lower recall either, due-to the trade-off
between them, therefore making our attack even more effective.}

The performance of the classifiers is shown in Table 1. The accuracy
was measured on the test set, which contains 36,000 samples.

\begin{table}[htbp]
\caption{Classifier Performance}
\centering{}%
\begin{tabular}{|c|c||c|c|}
\hline 
Classifier Type & Accuracy (\%) & Classifier Type & Accuracy (\%)\tabularnewline
\hline 
\hline 
RNN & 97.90 & Bidirectional GRU & 98.04\tabularnewline
\hline 
BRNN & 95.58 & Fully-Connected DNN & 94.70\tabularnewline
\hline 
LSTM & 98.26 & 1D CNN & 96.42\tabularnewline
\hline 
Deep LSTM & 97.90 & Random Forest & 98.90\tabularnewline
\hline 
BLSTM & 97.90 & SVM & 86.18\tabularnewline
\hline 
Deep BLSTM & 98.02 & Logistic Regression & 89.22\tabularnewline
\hline 
GRU & 97.32 & Gradient Boosted Decision Tree & 91.10\tabularnewline
\hline 
\end{tabular}
\end{table}

As can be seen in Table 1, the LSTM variants are the best malware
classifiers, accuracy-wise, and, as shown in Table 2, BLSTM is also
one of the classifiers most resistant to the proposed attack.

.

\subsection{Attack Performance}

In order to measure the performance of an attack, we consider two
factors:

The \emph{attack effectiveness} is the number of malware samples in
the test set which were detected by the target classifier, for which
the adversarial sequences generated by Algorithm 2 were misclassified
by the target malware classifier. 

\begin{equation}
\mathit{attack\_effectiveness}=\frac{|\{f(\boldsymbol{\mathbf{x}})=Malicious\lor f(\boldsymbol{\mathbf{x}}^{*})=Benign\}|}{|\{f(\boldsymbol{\mathbf{x}})=Malicious\}|}
\end{equation}

\[
s.t.\;\boldsymbol{\mathbf{x}}\in TestSet(f),\hat{f_{T}}=Algorithm1(f,T,X_{1},\epsilon),
\]

\[
\boldsymbol{\mathbf{x}}^{*}=Algorithm2(f,\hat{f_{T}},\boldsymbol{\mathbf{x}},n,D)
\]

We also consider the overhead incurred as a result of the proposed
attack. The \emph{attack overhead} is the average percentage of the
number of API calls which were added by Algorithm 2 to a malware sample
successfully detected by the target classifier, in order to make the
modified sample classified as benign (therefore calculated only for
successful attacks) by the black-box model:

\begin{equation}
\mathit{attack\_overhead}=avg(\frac{added\_APIs}{l})
\end{equation}

The average length of the API call sequence is: $avg(l)\approx100,000$.
The adversary chooses the architecture for the surrogate model without
any knowledge of the target model's architecture. We chose a GRU surrogate
model with 64 units (different from the malware classifiers used in
Section 4.2), which has a shorter training time compared to other
RNN variants, e.g., LSTM, which provides similar attack effectiveness.
Besides the classifier's type and architecture, we also used a different
optimizer for the surrogate model (ADADELTA instead of Adam). In our
implementation, we used the \href{https://github.com/tensorflow/cleverhans}{CleverHans library}.

Based on Equations 4 and 5, the proposed attack's performance is specified
in Table 2 (average of three runs).

\begin{table}[htbp]
\caption{Attack Performance}
\centering{}%
\begin{tabular}{|>{\centering}p{0.15\paperwidth}|>{\centering}p{0.1\paperwidth}|>{\centering}p{0.07\paperwidth}||>{\centering}p{0.1\paperwidth}|>{\centering}p{0.07\paperwidth}|>{\centering}p{0.07\paperwidth}|}
\hline 
Classifier Type & Attack Effectiveness (\%) & Additional API Calls (\%) & Classifier Type & Attack Effectiveness (\%) & Additional API Calls (\%)\tabularnewline
\hline 
\hline 
RNN & 100.0 & 0.0023 & Bidirectional GRU & 95.33 & 0.0023\tabularnewline
\hline 
\hline 
BRNN & 99.90 & 0.0017 & Fully-Connected DNN & 95.66 & 0.0049\tabularnewline
\hline 
\hline 
LSTM & 99.99 & 0.0017 & 1D CNN & 100.0 & 0.0005\tabularnewline
\hline 
\hline 
Deep LSTM & 99.31 & 0.0029 & Random Forest & 99.44 & 0.0009\tabularnewline
\hline 
\hline 
BLSTM & 93.48 & 0.0029 & SVM & 70.90 & 0.0007\tabularnewline
\hline 
\hline 
Deep BLSTM & 96.26 & 0.0041 & Logistic Regression & 69.73 & 0.0007\tabularnewline
\hline 
\hline 
GRU & 100.0 & 0.0016 & Gradient Boosted Tree & 71.45 & 0.0027\tabularnewline
\hline 
\end{tabular}
\end{table}

We can see in Table 2 that the proposed attack has very high effectiveness
and low attack overhead against all of the tested malware classifiers.
The attack effectiveness is lower for traditional machine learning
algorithms, such as SVM, due to the greater difference between the
decision boundaries of the GRU surrogate model and the target classifier.
Randomly modifying APIs resulted in significantly lower effectiveness
for all classifiers (e.g., 50.29\% for fully-connected DNN).

As mentioned in Section 4.1, $|TestSet(f)|=36,000$ samples, and the
test set $TestSet(f)$ is balanced, so the attack performance was
measured on: $|\{f(\boldsymbol{\mathbf{x}})=Malicious|\boldsymbol{\mathbf{x}}\in TestSet(f)\}|=18,000$
samples. For the surrogate model we used a perturbation factor of
$\epsilon=0.2$ and a learning rate of 0.1. $|X_{1}|=70$ samples
were randomly selected from the test set of 36,000 samples. We used
$T=6$ surrogate epochs. Thus, as shown in Section 3.2, the training
set size for the surrogate model is: $|X_{6}|=2^{5}*70=2240$ samples;
only 70 ($=|X_{1}|$) of the samples were selected from the test set
distribution, and all of the others were synthetically generated.
Using lower values, e.g., $|X_{1}|=50$ or $T=5$, achieved worse
attack performance, while larger values do not improve the attack
performance and result in a longer training time. The 70 samples from
the test set don't cover all of the malware families in the training
set; the effectiveness of the surrogate model is due to the synthetic
data.

For simplicity and training time, we used $m=n$ for Algorithm 2,
i.e., the sliding window size of the adversary is the same as that
used by the black-box classifier. However, even if this is not the
case, the attack effectiveness isn't degraded significantly. If $n>m$,
the adversary would keep trying to modify different API calls' positions
in Algorithm 2, until he/she modifies the ones impacting the black-box
classifier as well, thereby increasing the attack overhead without
affecting the attack effectiveness. If $n<m$, the adversary can modify
only a subset of the API calls affecting the black-box classification,
and this subset might not be diverse enough to affect the classification
as desired, thereby reducing the attack effectiveness. The closer
$n$ and $m$ are, the better the attack performance. For $n=100,m=140$,
there is an average decrease of attack effectiveness from 99.99\%
to 99.98\% for a LSTM classifier.

\subsubsection{Comparison to Previous Work}

Besides \cite{Hu17b} which was written concurrently and independently
from our work, \cite{Papernot16d} is the only recently published
RNN adversarial attack. The differences between that attack and the
attack addressed in this paper are mentioned in Section 2. We compared
the attacks in terms of performance. The attack effectiveness for
the IMDB dataset was the same (100\%), but our attack overhead was
better: 11.25 added words per review (on average), instead of 51.25
words using the method mentioned in \cite{Papernot16d}.

\subsection{Transferability for RNN Models}

While transferability was covered in the past in the context of DNNs
(e.g., \cite{Szegedy14}), to the best of our knowledge, this is the
first time it is \emph{evaluated} in the context of RNNs, proving
that the proposed attack is generic, not just effective against a
specific RNN variant, but is also transferable between RNN variants
(like LSTM, GRU, etc.), feed forward DNNs (including CNNs), and even
traditional machine learning classifiers such as SVM and random forest.

Two kinds of transferability are relevant to this paper: 1) the adversary
can craft adversarial examples against a surrogate model with a different
architecture and hyper parameters than the target model, and the same
adversarial example would work against both (\cite{Papernot16c}),
and 2) an adversarial example crafted against one target classifier
type might work against a different type of target classifier.

Both forms of transferability are evaluated as follows: 1) As mentioned
in Section 4.3, we used a GRU surrogate model. However, as can be
seen in Table 2, the attack effectiveness is high, even when the black-box
classifier is not GRU. Even when the black-box classifier is GRU,
the hyper parameters (such as the number of units and the optimizer)
are different. 2) The attack was designed against RNN variants; however,
we tested it and found the attack to be effective against both feed
forward networks and traditional machine learning classifiers, as
can be seen in the last six lines of Table 2. Our attack is therefore
effective against all malware classifiers.

\section{GADGET: End-to-End Attack Framework Description}

To verify that an attacker can create an end-to-end attack using the
proposed method (Section 3), we implemented \textbf{GADGET}: \textbf{G}enerative
\textbf{A}pi a\textbf{D}versarial \textbf{G}eneric \textbf{E}xample
by \textbf{T}ransferability framework. This is an end-to-end attack
generation framework that gets a black-box classifier ($f$ in Section
3) as an input, an initial surrogate model training set ($X_{1}$
in Algorithm 1), and a malware binary to evade $f$, and outputs a
modified malware binary whose API call sequence is misclassified by
$f$ as benign, generating the surrogate model ($\hat{f}$ in Algorithm
1) in the process.

GADGET contains the following components: 1) Algorithms 1 and 2, implemented
in Python, using Keras with TensorFlow back end, 2) A C++ Wrapper
to wrap the malware binary and modify its generated API call sequence
during run time, and 3) A Python script that wraps the malware binary
with the above mentioned wrapper, making it ready to deploy. The components
appear in Figure 3.

\begin{figure}[htbp]
\begin{centering}
\subfloat[Malware Binary, without GADGET]{\includegraphics[scale=0.35]{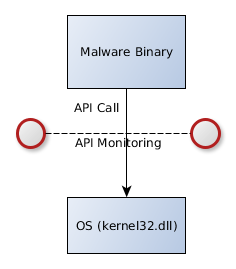}}\hfill{}\subfloat[Malware Binary, with GADGET]{\includegraphics[scale=0.35]{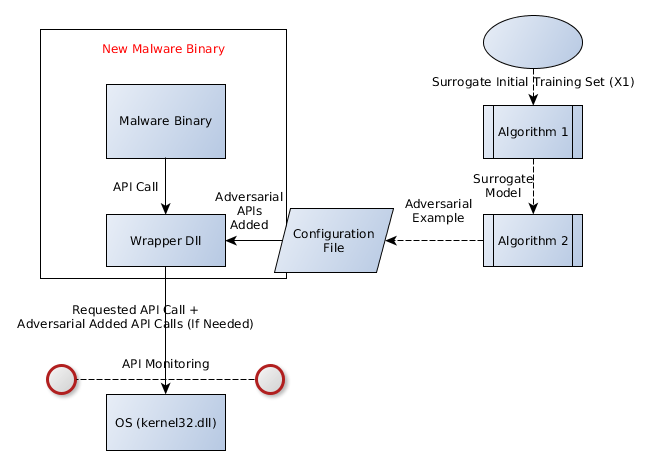}}
\par\end{centering}
\caption{Malware Binary, with and without GADGET}
\end{figure}

\subsubsection{Adding API Calls without Damaging Functionality}

As mentioned in Section 3.2, we implemented Algorithm 2 using a \emph{mimicry
attack} \cite{Wagner02}. We discarded \emph{equivalence attacks}
and \emph{disguise attacks} (Section 2), since they lack the flexibility
needed to modify \emph{every API call}, and thus are not robust enough
to camouflage every malware. Therefore, we implemented a \emph{no-op
attack,} adding APIs which would have no effect on the code's functionality.
Since some API call monitors (such as Cuckoo Sandbox) also monitor
the return value of an API call and might ignore failed API calls,
we decided to implement the API addition by adding no-op API calls
with valid parameters, e.g., reading 0 bytes from a valid file. This
was more challenging to implement than calling APIs with invalid arguments
(e.g., reading from an invalid file handle), since a different implementation
should be used for each API. However, this effort can be done once
and can subsequently be used for every malware, as we've done in our
framework. This makes detecting those no-op APIs much harder, since
the API call runs correctly, with a return value indicative of success.
The functionality validation of the modified malware is discussed
in Section 5.1. Further measures, such as randomized arguments, can
be taken by the attacker to prevent the detection of the no-op APIs
by analyzing the arguments of the API calls. Attacking a classifier
with argument inputs is discussed in Section 5.2.

\subsubsection{Implementing a Generic Framework}

The requirements for the generic framework are: 1) there is no access
to the malware source code (access only to the malware binary executable),
and 2) the same code should work for every adversarial sample: no
adversarial example-specific code should be written. The reasons for
these requirements are two-fold. First, adding the code as a wrapper,
without changing the malware's business logic makes the framework
more robust to modification of the malware classifier model, preventing
another session of malware code modification and testing. Second,
with the Malware-as-a-Service trend, not everyone who uses a malware
has its code. Some ransomwares are automatically generated using minimal
configuration (e.g., only the CNC server is modified by the user),
without source code access. Thus, the GADGET framework expands the
number of users that can produce an evasive malware from malware developers
to every person that purchases a malware binary, making the threat
much greater.

In order to meet those requirements, we wrap the malware binary from
the outside with proxy code between the malware code and the OS DLLs
implementing the API calls (e.g., kernel32.dll), fulfilling requirement
\#1. The wrapper code gets the adversarial sequence for the malware
binary, generated by Algorithm 2, as a configuration file. The logic
of this wrapper code is to hook all APIs that will be monitored by
the malware classifier. These API calls are known to the attacker,
as mentioned in Section 3.2. These hooks call the original APIs (to
preserve the original malware functionality), keep track of the API
sequence executed so far, and call the adversarial example's additional
APIs in the proper position based on the configuration file (so they
will be monitored by the malware classifier), instead of hard-coding
the adversarial sequence to the code (fulfilling requirement \#2).
This flow is presented in Figure 3(b).

We generated a new malware binary that contains the wrapper's hooks
by patching the malware binary's IAT using \href{http://hasherezade.github.io/IAT_patcher/}{IAT Patcher},
redirecting the IAT's API calls' addresses to the matching C++ wrapper
API hook implementation. That way, if another hook (e.g., Cuckoo Sandbox)
monitors the API calls, the adversarial APIs are already being called
and monitored like any regular API call. To affect dynamic libraries,
\emph{LdrGetProcedureAddress()\textbackslash{}GetProcAddress}() hook
has additional functionality: it doesn't return a pointer to the requested
procedure, but instead returns a pointer to a wrapper function that
implements the previously described regular static hook functionality
around the requested procedure (e.g., returning a pointer to a wrapper
around \emph{WriteFile()} if ``WriteFile'' is the argument to \emph{GetProcAddress}()).
When the malware code calls the pointer, the hook functionality will
be called, transparent to the user.

The code is POC and does not cover all corner cases, e.g., wrapping
a packed malware, which requires special handling for the IAT patching
to work, or packing the wrapper code to evade statically signing it
as malicious (its functionality is implemented inline, without external
API calls, so dynamic analysis of it is challenging). We avoided running
Algorithm 2 inside the wrapper, and used the configuration file to
store the modified APIs instead, thus preventing much greater overhead
for the (wrapped) malware code.

\subsection{Adversarial Example Functionality Validation}

In order to \emph{automatically} verify that we do not harm the functionality
of the malware we modify, we monitored each sample in Cuckoo Monitor
before and after the modification. We define the modified sample as
\emph{functionality preserving} if the API call sequence after the
modification is the same as before the modification when comparing
API type, return value and order of API calls, except for the added
API calls, which return value should always be a success value. We
found that all of the 18,000 modified samples are \emph{functionality
preserving.}

One of the families that did not exist in the training set was the
WannaCry ransomware. This makes it an excellent candidate to \emph{manually}
analyze GADGET's output. First, we ran the sample via Cuckoo Sandbox
and recorded its API calls. The LSTM malware classifier mentioned
in Section 4.2 successfully detected it as malicious, although it
was not part of the training set. Then we used GADGET to generate
a new WannaCry variant, providing this variant the configuration file
containing the adversarial sequence generated by Algorithm 2. We ran
the modified WannaCry binary, wrapped with our framework and the configuration
file, in Cuckoo Sandbox again, and fed the recorded API call sequence
to the same LSTM malware classifier. This time, the malware classifier
classification was benign, although the malicious functionality remains:
files were still being encrypted by the new binary, as can be seen
in the Cuckoo Sandbox snapshot and API call sequence. This means that
the proposed attack was successful, end-to-end, without damaging WannaCry's
functionality.

\subsection{Handling API Arguments}

We now modify our attack to evade classifiers that analyze arguments
as well. In order to represent the API call arguments, we used MIST
\cite{Trinius09}, as was done by other malware classifiers, e.g.,
MALHEUR \cite{Rieck11}. MIST (Malware Instruction Set) is a representation
for monitored behavior of malicious software, optimized for analysis
of behavior using machine learning. Each API call translates to an
instruction. Each instruction has levels of information. The first
level corresponds to the category and name of a monitored API call.
The following levels of the instruction contain different blocks of
arguments. The main idea underlying this arrangement is to move \textquotedblleft noisy\textquotedblright{}
elements, such as the loading address of a DLL, to the end of an instruction,
while discriminative patterns, such as the loaded DLL file path, are
kept at the beginning of the instruction. We used MIST level 2. We
converted our Cuckoo Sandbox reports to MIST using \href{https://github.com/M-Gregoire/Cuckoo2Mist}{Cuckoo2Mist}.
We extracted a total of 220 million lines of MIST instructions from
our dataset. Of those, only several hundred of lines were unique,
i.e., different permutations of argument values extracted in MIST
level 2. This means that most API calls differ only in arguments that
are not relevant to the classification or use the same arguments.
To handle MIST arguments, we modified our attack in the following
way: Instead of one-hot encoding every API call type, we one-hot encoded
every unique {[}API call type, MIST level 2 arguments{]} combination.
Thus, \emph{LoadLibrary}(``kernel32.dll'') and \emph{LoadLibrary}(``user32.dll'')
are now regarded as separate APIs by the classifier. Our framework
remains the same, where Algorithm 2 selects the most impactful combination
instead of API type. However, instead of adding combinations that
might harm the code's functionality (e.g., \emph{ExitWindowsEx}()),
we simply add a different API call type (the one with the minimal
Jacobian value) in Algorithm 2, which would \textbf{not} cause this
issue. We now assume a more informed attacker, who knows not just
the exact encoding of each API type, but also the exact encoding of
every argument combination. This is a reasonable assumption since
arguments used by benign programs, like Windows DLLs file paths, are
known to attackers \cite{Huang11}.

Handling other API arguments (and not MIST level 2) would be similar,
but require more preprocessing (word embedding, etc.) with a negligible
effect on the classifier accuracy. Thus, focusing only on the most
important arguments (MIST level 2) that can be used by the classifier
to distinguish between malware and benign software, as done in other
papers (\cite{Huang16}), proves that analyzing arguments is not an
obstacle for the proposed attack.

\subsection{Handling Hybrid Classifiers and Multiple Feature Types}

Since our attack modifies only a specific feature type (API calls),
combining several types of features might make the classifier more
resistant to adversarial examples against a specific feature type.
Some real-world next generation anti-malware products (such as SentinelOne)
are hybrid classifiers, combining both static and dynamic features
for a better detection rate.

Our attack can be extended to handle hybrid classifiers using two
phases: 1) the creation of a \emph{combined surrogate} model, including
all features, using Algorithm 1, and 2) attacking \emph{each feature
type in turn} with a specialized attack, using the surrogate model.
If the attack against a feature type fails, we continue and attack
the next feature type until a benign classification by the target
model is achieved or until all feature types have been (unsuccessfully)
attacked.

We decided to use printable strings inside a PE file as our static
features, as they are commonly used as the static features of state
of the art hybrid malware classifiers \cite{Huang16}, although any
other modifiable feature type can be used. Strings can be used, e.g.,
to statically identify loaded DLLs and called functions, recognize
modified file paths and registry keys, etc. Our architecture for the
hybrid classifier, shown in Figure 2(b), is: 1) A dynamic branch that
contains an input vector of 140 API calls, each one-hot encoded, inserted
into a LSTM layer of 128 units, and sigmoid activation function, with
a dropout rate of 0.2 for both inputs and recurrent states. 2) A static
branch that contains an input vector of 20,000 Boolean values: for
each of the 20,000 most frequent strings in the entire dataset, do
they appear in the file or not? (analogous to a similar procedure
used in NLP, which filters the least frequent words in a language).
This vector is inserted into two fully-connected layers with 128 neurons,
a ReLU activation function, and a dropout rate of 0.2 each. The 256
outputs of both branches are inserted into a fully-connected output
layer with sigmoid activation function. Therefore, the input of the
classifier is a vector containing 140 one-hot encoded APIs and 20,000
Boolean values, and the output is malicious or benign classification.
All other hyper parameters are the same as in Section 4.2. The surrogate
model used has a similar architecture to the attacked hybrid model
described above, but it uses a different architecture and hyper parameters:
GRU instead of LSTM in the dynamic branch and 64 hidden units instead
of 128 in both static and dynamic surrogate branches. Due to hardware
limitations, we used just a subset of the dataset: 54,000 training
samples and test and validation sets of 6,000 samples each. The dataset
was representative and maintained the same distribution as the dataset
described in Section 4.1. Trained on this dataset, a classifier using
only the dynamic branch (Figure 2(a)) reaches 92.48\% accuracy on
the test set, a classifier using only the static branch attains 96.19\%
accuracy, and a hybrid model, using both branches (Figure 2(b)) achieves
96.94\% accuracy, meaning that using multiple feature types improves
the accuracy.

We used two specialized attacks: an attack against API call sequences
and an attack against printable strings. The API sequence attack is
Algorithm 2. When performing it against the hybrid classifier, without
modifying the static features of the sample, the attack effectiveness
(Equation 4) decreases to 45.95\%, compared to 96.03\% against a classifier
trained only on the dynamic features, meaning that the attack was
mitigated by the use of additional features. The strings attack is
a variant of the attack described in \cite{Grosse16}, using the surrogate
model instead of the attacked model used in \cite{Grosse16} to compute
the gradients in order to select the string to add, while the adversarial
sample's maliciousness is still tested against the attacked model,
making this method a black-box attack. In this case, the attack effectiveness
is 68.66\%, compared to 77.33\% against a classifier trained only
on the static features. Finally, the combined attack's effectiveness
against the hybrid model was 82.27\%. Other classifier types provide
similar results which are not presented here due to space limits.

We designed GADGET with the ability to handle a hybrid model, by adding
its configuration file's static features' modification entries. Each
such string is appended to the original binary before being IAT patched,
either to the EOF or to a new section, where those modifications don't
affect the binary's functionality.

To summarize, we have shown that while the usage of hybrid models
decreases the specialized attacks' effectiveness, using our suggested
hybrid attack achieves high effectiveness. While not shown due to
space limits, the attack overhead isn't significantly affected.

\section{Conclusions and Future Work}

In this paper, we demonstrated a generic black-box attack, generating
adversarial sequences against API call sequence based malware classifiers.
Unlike previous adversarial attacks, we have shown an attack with
a verified effectiveness against all relevant common classifiers:
RNN variants, feed forward networks, and traditional machine learning
classifiers. Therefore, this is a true black-box attack, which requires
no knowledge about the classifier besides the monitored APIs. We also
created the GADGET framework, showing that the generation of the adversarial
sequences can be done end-to-end, in a generic way, without access
to the malware source code. Finally, we showed that the attack is
effective, even when arguments are analyzed or multiple feature types
are used. Our attack is \emph{the first practical }\textbf{\emph{end-to-end}}\emph{
attack} dealing with all of the subtleties of the cyber security domain,
posing a concrete threat to next generation anti-malware products,
which have become more and more popular. While this paper focus on
API calls and printable strings as features, the proposed attack is
valid for every modifiable feature type, static or dynamic.

Our future work will focus on two areas: defense mechanisms against
such attacks and attack modifications to cope with such mechanisms.
Due to space limits, we plan to publish an in depth analysis of various
defense mechanisms in future work. The defense mechanisms against
such attacks can be divided into two subgroups: 1) detection of adversarial
examples, and 2) making the classifier resistant to adversarial attacks.
To the best of our knowledge, there is currently no published and
evaluated method to detect or mitigate RNN adversarial sequences.
This will be part of our future work. We would also compare between
the effectiveness of different surrogate models' architecture.


\begin{thebibliography}{10}
\bibitem{Arjovsky17}M. Arjovsky and L. Bottou. Towards Principled
Methods for Training Generative Adversarial Networks. In ICLR, 2017.

\bibitem{Carlini17}N. Carlini and D. Wagner, Towards Evaluating the
Robustness of Neural Networks, In IEEE S\&P, 2017.

\bibitem{Chen17}P. Y. Chen, H. Zhang, Y. Sharma, J. Yi, and C. J.
Hsieh. Zoo: Zeroth order optimization based black-box attacks to deep
neural networks without training substitute models. In ACM Workshop
on Artificial Intelligence and Security, 2017.

\bibitem{Goodfellow15}I. J. Goodfellow, J. Shlens, and C. Szegedy.
Explaining and Harnessing Adversarial Examples. In ICLR, 2015.

\bibitem{Grosse16}K. Grosse, N. Papernot, P. Manoharan, M. Backes,
and P. McDaniel. Adversarial examples for malware detection. In ESORICS,
2017.

\bibitem{Hu17a}W. Hu and Y. Tan. Generating adversarial malware examples
for black-box attacks based on GAN. ArXiv e-prints, abs/1702.05983,
2017.

\bibitem{Hu17b}W. Hu and Y. Tan. Black-box attacks against RNN based
malware detection algo- rithms. ArXiv e-prints, abs/1705.08131, 2017.

\bibitem{Huang11}L. Huang, A. D. Joseph, B. Nelson, B. I.P. Rubinstein,
and J. D. Tygar. Adversarial machine learning. In ACM Workshop on
Security and Artificial Intelligence, 2011.

\bibitem{Huang16}W. Huang and J. W. Stokes. MtNet: A multi-task neural
network for dynamic malware classification. In DIMVA, 2016.

\bibitem{Papernot16b}N. Papernot, P. McDaniel, S.h Jha, M. Fredrikson,
Z. B. Celik, and A. Swami. The limitations of deep learning in adversarial
settings. In IEEE European Symposium on Security and Privacy, 2016.

\bibitem{Papernot16c}N. Papernot, P. McDaniel, I. Goodfellow, S.
Jha, Z. B. Celik, and A. Swami. Practical black-box attacks against
machine learning. In ASIACCS, 2017.

\bibitem{Papernot16d}N. Papernot, P. McDaniel, A. Swami, and R. Harang.
Crafting adversarial input sequences for recurrent neural networks.
In IEEE MILCOM, 2016.

\bibitem{Pascanu15}R. Pascanu, J. W. Stokes, H. Sanossian, M. Marinescu,
and A. Thomas. Malware classification with recurrent networks. In
IEEE ICASSP, 2015.

\bibitem{Rieck11}K. Rieck, P. Trinius, C. Willems, and T. Holz. Automatic
analysis of malware behavior using machine learning. In Journal of
Computer Security, 2011.

\bibitem{Rosenberg16}I. Rosenberg and E. Gudes. Bypassing system
calls-based intrusion detection systems. In Concurrency and Computation:
Practice and Experience, 2016.

\bibitem{Rosenberg17}I. Rosenberg, G. Sicard, and E. David. DeepAPT:
Nation-state APT attribution using end-to-end deep neural networks.
In International Conference of Artificial Neural Networks, 2017.

\bibitem{Rosenberg17b}I. Rosenberg, A. Shabtai, L. Rokach, Y. Elovici.
Low Resource Black-Box End-to-End Attack Against State of the Art
API Call Based Malware Classifiers, arXiv preprint arXiv:1804.08778,
2018.

\bibitem{Szegedy14}C. Szegedy, W. Zaremba, I. Sutskever, J. Bruna,
D. Erhan, I. J. Goodfellow, and R. Fergus. Intriguing properties of
neural networks. In ICLR, 2014.

\bibitem{Tandon06}G. Tandon and P. K. Chan. On the learning of system
call attributes for host-based anomaly detection. In International
Journal on Artificial Intelligence Tools, 2006.

\bibitem{Trinius09}P. Trinius, C. Willems, T. Holz, and K. Rieck.
A malware instruction set for behavior-based analysis. In Sicherheit,
2010.

\bibitem{Wagner02}D. Wagner and P. Soto. Mimicry attacks on host-based
intrusion detection systems. In ACM CCS, 2002.
\end{thebibliography}
\end{document}